\documentclass[twocolumn,pra,floatfix]{revtex4-1}
\usepackage[pdftex]{graphicx}
\usepackage{amsfonts,amsmath,amssymb}
\usepackage{bm}
\usepackage{dsfont}
\usepackage{color}
\usepackage[colorlinks=true,linkcolor=blue,urlcolor=blue,citecolor=blue]{hyperref}
\usepackage{times}
\usepackage{textcomp}

\newcommand{\td}{\dot\theta}
\newcommand{\ham}{\mathcal{H}}
\begin{document}
 
\title{Wavepacket dynamics on Chern band lattices in a trap}

\author{Sthitadhi Roy}
\author{Adolfo G. Grushin}
\author{Roderich Moessner}
\author{Masudul Haque}

\affiliation{Max-Planck-Institut f\"ur Physik komplexer Systeme, N\"othnitzer Stra{\ss}e 38, 01187
  Dresden, Germany}

\begin{abstract}

The experimental realization of lattices with Chern bands in ultracold-atom and  photonic systems
has motivated the study of time-dependent phenomena, such as spatial propagation, in lattices with
nontrivial topology.  We study the dynamics of gaussian wavepackets on the Haldane honeycomb
Chern-band lattice model, in the presence of a harmonic trap.  We focus on the transverse response
to a force, which is due partly to the Berry curvature and partly to the transverse component of the
energy band curvature.  We evaluate the accuracy of a semiclassical description, which treats the
wavepacket as a point particle in both real and momentum space, in reproducing the motion of a
realistic wavepacket with finite extent.  We find that, in order to accurately capture the
wavepacket dynamics, the extent of the wavepacket in momentum space needs to be taken into account:
The dynamics is sensitive to the interplay of band dispersion and Berry curvature over the
finite region of momentum (reciprocal) space where the wavepacket has support.  Moreover, if the
wavepacket is prepared with a finite initial momentum, the semiclassical analysis reproduces its
motion as long as it has a large overlap with the eigenstates of a single band.  The semiclassical
description generally improves with increasing real-space size of the wavepacket, as long as the
external conditions (e.g., external force) remain uniform throughout the spatial extent of the
wavepacket.

\end{abstract} 



\maketitle

\section{Introduction \label{sec:intro}}

The presence of non-trivial topology in the energy bands of lattice models \cite{Hasan2010,Qi2011}
has generated intense interest over the past decade.  When an energy band possessing a nonzero Chern
number is filled with fermions, one obtains a Chern insulator, realizing a quantized Hall
effect without external magnetic fields.  The Chern number is a topological invariant, defined as the
flux of the \emph{Berry curvature} over the Brillouin zone.  The Berry curvature acts like a
momentum space analogue of the magnetic field~\cite{Haldane04}.
One of the first examples of a Chern band model was introduced in the seminal work by
Haldane~\cite{Haldane88}, where time reversal symmetry is explicitly broken in a honeycomb lattice
model of free fermions by complex hoppings between next-nearest-neighbor lattice sites.

Chern band physics has been discussed in numerous different scenarios.  Originally, the focus was on
electronic systems~\cite{Kane2005a,Bernevig2006,Konig2007,Konig2008,Chang2013}.  More recently,
lattices with nontrivial Berry curvature have been experimentally realized using ultracold atoms
trapped in optical lattices~\cite{Aidelsburger2013,Miyake2013,Uehlinger2013b,Jotzu2014}.  The spatial
geometries in these setups are more flexible than in traditional solid-state situations.  These
experimental developments have thus motivated the study of effects of lattice and trap geometries on
topological states~\cite{Goldman2010,Liu2010,Stanescu2010,Buchhold2012,Deng2014, Kolovsky2014}.
Cold atoms also provide an excellent platform for observing and analyzing non-equilibrium dynamics
of Chern bands.  Therefore, there is increasing interest in developing protocols to characterize the
topological nature of Chern bands from non-equilibrium behavior
\cite{Killi2012,Killi2012a,Dauphin2013,Hauke2014, DAlessio2014, Caio2015, Goldman2013,
  Aidelsburger2015,S14, Price2012,GRH15}.  For instance, the quench dynamics of Chern insulators can signal
their non-trivial edge states~\cite{Goldman2013} and the semiclassical trajectory of wavepackets can
be related to the Chern number~\cite{Dauphin2013}.

Topological photonics has proved itself to be another important context for Chern band physics
\cite{Lu2014,Haldane2008,Raghu2008,Koch2010,Hafezi2011,Petrescu2012,Wang2009,Fang2012,Rechtsman2013,Hafezi2013}.
Theoretical studies and proposals \cite{Haldane2008,Raghu2008,Koch2010,Hafezi2011,Petrescu2012} have
also been supplemented by experimental observation of the topological edge states in photonic
systems~\cite{Wang2009,Fang2012,Rechtsman2013,Hafezi2013}. In addition, topological magnons have been proposed 
as a platform to study transport influenced by Berry curvature in the context of the thermal Hall effect \cite{Matsumoto2011,Matsumoto2011B,Shindou2013}.

For ultracold atoms (especially bosons) and for photonic systems, a dynamical situation where the
atoms/photons form a spatially localized and evolving wavepacket is more natural than a static
situation in which a band is exactly filled.  Thus, the recent experimental developments call for a
systematic understanding of the non-equilibrium dynamics of realistic wavepackets on Chern band
models.  
A particularly important theme is the response of a localized wavepacket to an applied force
(potential gradient) \cite{Price2012,Dauphin2013,Jotzu2014, Aidelsburger2015,Duca2015}.  One expects
Bloch oscillations in the direction of the applied force.
In addition, there is also motion perpendicular to the direction of the force, with contributions
due to the topological Berry curvature and due to the band dispersion.

A natural framework to describe the motion of wavepackets is to use semiclassical equations of
motion. For systems with Berry curvature, the semiclassical equations were introduced in the context
of crystals with magnetic Bloch
bands~\cite{Chang1995,Chang1996,Sundaram1999,Jungwirth2002,Gosselin2006, Xiao2010},
anomalous Hall responses \cite{Xiao2006, Xiao2010, Sinitsyn2008}, 
and optical lattices with spin-orbit coupling~\cite{Dudarev2004}. They have been used to study the
effect of the Berry curvature on wavepacket trajectories~\cite{Diener2003,Price2012} and on
collective modes~\cite{vanderBijl2011, Price2013} in ultracold gases. For instance,
Ref.~\cite{Price2012} outlines a procedure to isolate the contribution of the topological Berry
curvature by separately evolving the system under opposite potential gradients and then
appropriately summing up the responses.

Semiclassical approaches typically rely on the approximation of assigning a sharply defined position
and momentum simultaneously to a quantum state.
Based on this assumption one can solve the set of coupled differential equations for position and
momenta and hence obtain sharply defined trajectories of the particle in both real and momentum
space. However, this assumption is \emph{a priori} not valid in realistic situations where the
wavepackets have a finite spread in both real and momentum space, a scenario expected in typical
ultracold bosonic and photonic experiments.

In this work, we study the effect of size and initial momentum of a wavepacket placed off-center in
a harmonic trap in Haldane's honeycomb model, and evaluate the ability of the semiclassical approach
to describe these effects.  We focus on time evolutions up to moderate timescales such that the
displacement of the wavepacket from its initial position is not large compared to the wavepacket
size and much smaller than the distance to the center of the trap.  In this regime, the trap plays a
role similar to a constant force.
We quantify the transverse motion of the wavepacket using the angular velocity $\dot{\theta}$ with
respect to the center of the trap.

We find that the standard point-particle semiclassical approach captures some qualitative features
of the dynamics, but is generally insufficient to quantitatively describe the actual real time
dynamics.  Therefore, we reformulate the semiclassical description to take into account the finite
spread of the wavepacket in momentum space.  The exact evolution of $\dot{\theta}$ is compared in
detail to predictions from the extended semiclassics and from the standard point-particle
semiclassics.
In momentum space, the wavepacket moves at a constant rate in the direction of the force, leading to Bloch
oscillations.  As it crosses different regions of the Brillouin zone, the local Berry curvature and
band curvature determine $\dot\theta(t)$.  For spatially localized wavepackets, the extent in
momentum space is finite.  The ``extended semiclassics'' procedure incorporates the variations of
band dispersion and Berry curvature in this extended region of momentum space.  We find that, as
long as the physics is dominated by one band, this procedure reproduces the full dynamics well.
This shows that the basic idea of semiclassics (simultaneously assigning both position and momentum
to a quantum particle) can incorporate aspects of the full quantum dynamics to an extent beyond what is
known from the usual point-particle treatment.

One might intuitively expect that semiclassical descriptions should work better for spatially large
wavepackets, since these correspond to smaller regions in momentum space.
We show that this is generally true, but that semiclassics still describes the dynamics of rather
small wavepackets, especially if momentum-space extent is included.  In addition, by considering a
tight trap, we show an example of possible experimental relevance where larger real-space sizes can
render the semiclassical description less inaccurate, due to an inhomogeneity of the force within
the spatial support region of the wavepacket.

We also demonstrate the effect of initializing the wavepacket with a finite momentum.  In addition
to zero momentum ($\mathbf{\Gamma}$ point) we start with the packet at one of the $\mathbf{K}$
points and one of the $\mathbf{M}$ points of the Brillouin zone.  We show that imparting momenta to
a gaussian wavepacket using an $e^{i\mathbf{k}\cdot\mathbf{r}}$-like factor can cause the
wavefunction to have significant occupancy in the upper band, including extreme cases where it is
almost completely transferred to the upper band.  As long as one of the bands dominates, the
semiclassical description works well when using the properties (band dispersion and Berry curvature)
of the band where the state has most of its weight.  The single-band semiclassical procedure is
naturally insufficient when multiple bands are significantly occupied: Features like interference oscillations
are not captured by an incoherent averaging of contributions from different bands.

\begin{figure}
 \includegraphics[width=0.95\linewidth]{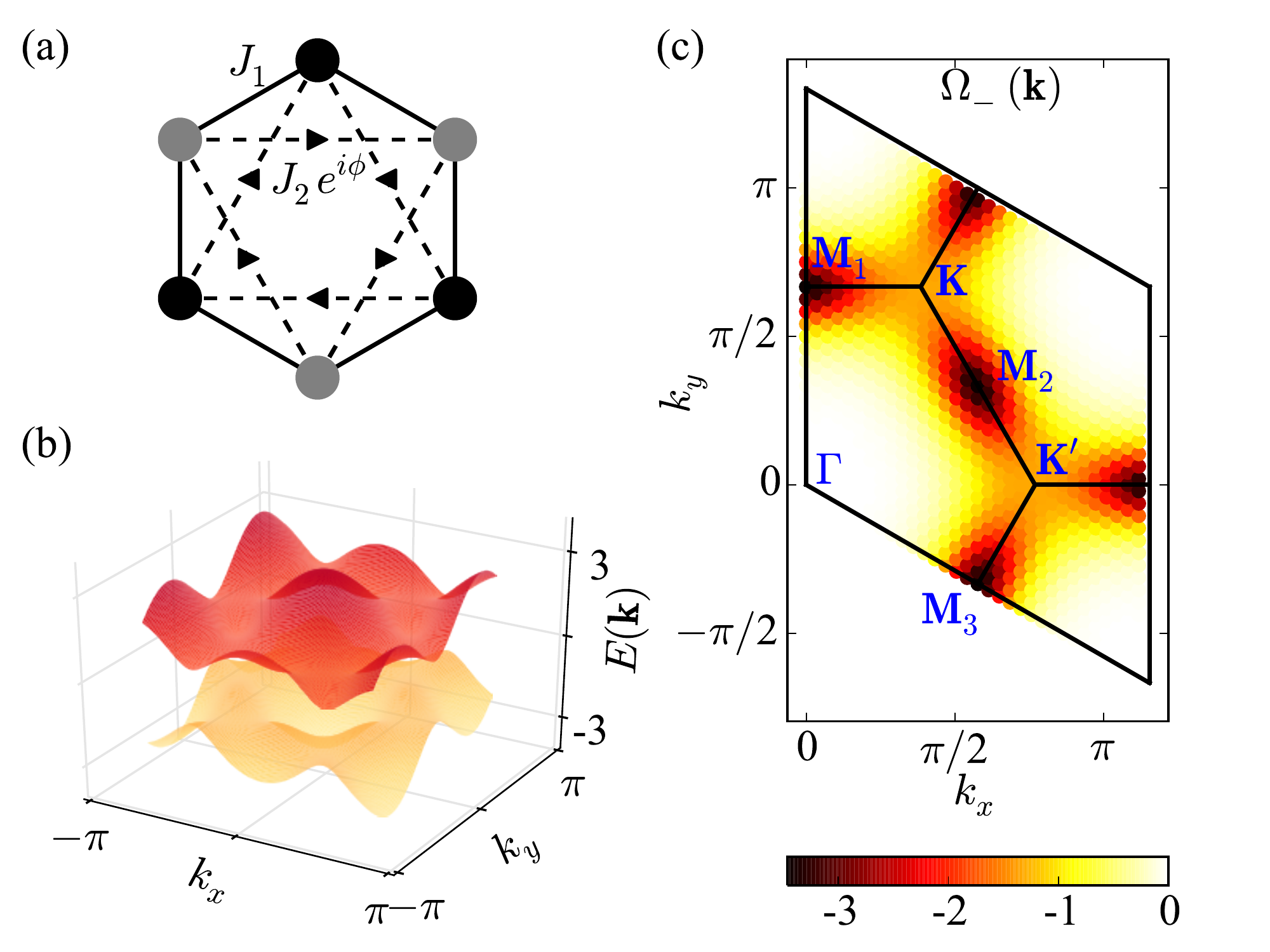}
 \caption{(Color online) (a) Honeycomb lattice defining the Haldane model \eqref{eq:model} with
   black and grey circles denoting the two sublattices. The solid lines represent the real
   nearest-neighbor hoppings whereas the dotted lines the complex next-nearest-neighbor
   hoppings. The arrows indicate the direction in which the particle hops to pick up a phase
   $\phi$. (b) The two band energy spectrum with parameters $J_1=-1$, $J_2=-1/4$, and
   $\phi=0.49\pi$. (c) The Berry curvature in the lower band for the same parameter values shown as
   a color map in the Brillouin zone.  The high symmetry points are marked. The
   boundaries of the Brillouin zone are formed by the reciprocal lattice vectors
   $\mathbf{G}_1=(0,4\pi/3)$ and $\mathbf{G}_2=(2\pi/\sqrt{3},-2\pi/3)$.  }
 \label{fig:model}
\end{figure}

The rest of the paper is organized as follows: in Sec.\ref{sec:model} we describe the model, the
geometries and the simulations of time evolution. In Sec.\ref{sec:semiclassics} we discuss the
semiclassical framework for the dynamics of the wavepacket, followed by a comparison of the results
from simulations and semiclassics in Sec.\ref{sec:thetadot}. Finally, we comment on the dynamics in
a tight harmonic trap in Sec.\ref{sec:tight} and provide some discussion and context in
Sec.\ref{sec:summary}.

\section{Simulations of wavepacket dynamics on the Haldane model \label{sec:model}}

\subsection{Model Hamiltonian}

Haldane's model~\cite{Haldane88} is a tight-binding Hamiltonian on a honeycomb lattice:
\begin{equation}
\label{eq:model}
 \ham_{\text{HM}} = J_1\sum_{\langle i,j\rangle}\hat{b}_i^\dagger\hat{b}_j+J_2\sum_{\langle\langle i,j\rangle\rangle}e^{i\phi_{ij}}\hat{b}_i^\dagger\hat{b}_j + \text{h.c}
\end{equation}
where the $\langle i,j \rangle$ denotes the nearest neighbors, $\langle\langle i,j\rangle\rangle$
denotes the next nearest neighbors and $\hat{b}^{\dagger}_j$ $(\hat{b}_j)$ is the creation
(annihilation) operator at site $i$.  We will formally consider the dynamics of a single particle, so $\hat{b}^{\dagger}_j$, $\hat{b}_j$
may be thought of as either fermionic or bosonic operators.  
When particles hop between next nearest neighbors they pick
up a phase $\phi_{ij}=\phi$ if they hop in the direction of the arrow shown in
Fig.~\ref{fig:model}(a) and $-\phi$ if they hop in the opposite direction.  The energy spectrum is
gapped if $\phi\neq n\pi$ and is particle hole symmetric for $\phi=\pi/2$.  Throughout
this paper we use the parameters $J_1=-1,$ $J_2=-1/4,$ and $\phi=0.49\pi$, for which the energy
spectrum is depicted in Fig.~\ref{fig:model}(b).  The parameters are chosen so that the two energy
bands are quite similar and none of them are excessively flat (since band flatness can introduce
additional peculiarities in the dynamics), and so as to avoid exact particle-hole symmetry, since we
are interested in generic rather than fine-tuned effects.  We expect the physics described in this
paper (explicitly for these parameters) to be exemplary for a wide region in parameter space.

For these parameters the bands have a finite Chern
number; $-1$ for the lower band and $+1$ for the upper band.  The distribution of the Berry
curvature of the lower band in the Brillouin zone is shown in Fig.~\ref{fig:model}(c).  The upper
band has approximately opposite Berry curvature, i.e., positive instead of negative values.  We provide
some details of the topological properties and conventions in Appendix \ref{sec:berrycurvature}.  
Fig.~\ref{fig:model}(c) also shows the high-symmetry points.  In addition to zero-momentum
($\mathbf{\Gamma}$ point), these are three inequivalent $\mathbf{M}$ points and two inequivalent
$\mathbf{K}$ points.  We choose the boundaries of the Brillouin zone to be a parallelogram.  In the
literature this is equivalently often chosen to be of hexagon shape; the solid lines inside the
Brillouin zone show the boundaries for such choice. 

We set $\hbar=1$, measuring time in units of $\hbar/J_1$ and energy in units of $J_1$. Space and
momentum are in units of lattice spacing (set to unity) and inverse lattice spacing, respectively and geometric angles are measured in radians.

\subsection{Construction of wavepackets \label{sec:construction}}

We are interested in the dynamics of a wavepacket of finite real-space extent in Haldane's
honeycomb model in the presence of a harmonic trap.  The initial wavepacket is prepared with
gaussian shape.  For zero initial momentum, 
\begin{equation} \label{eq:gaussian_wavepacket} 
\vert\psi(t=0)\rangle ~=~ \frac{1}{\mathcal{N}}\sum_l c_l\vert
  l\rangle ~=~ \frac{1}{\mathcal{N}}\sum_l e^{-\frac{\vert\mathbf{r}_l-\mathbf{r}_c\vert^2}{2\sigma^2}}\vert
  l\rangle.
\end{equation}
Here $\vert l \rangle$ denotes a single-particle state, with the particle completely localized at a site indexed by $l$ and
$\mathcal{N}$ is a normalization factor, and $\mathbf{r}_l=(x_l,y_l)$ denotes the Euclidean position
of site $l$.  The coefficient $c_l(t)$ denotes the amplitude of the wavefunction at time $t$ at
site $l$, and $\sigma$ is the width of the wavepacket.
We will also use wavepackets with nonzero initial momenta.  A finite momentum is obtained by
multiplying the coefficients with a site-dependent phase factor:
\begin{multline} \label{eq:initial_wavepacket_with_momentum}
  \vert\psi(t=0;\mathbf{k}_0)\rangle = \sum_l c_{l,0}e^{i \mathbf{k}_0\cdot \mathbf{r}_l}
\vert{l}\rangle
\\
= \frac{1}{\mathcal{N}}\sum_l \exp\left[-\frac{\vert\mathbf{r}_l-\mathbf{r}_c\vert^2}{2\sigma^2} + i\mathbf{k}_0\cdot
  \mathbf{r}_l \right] \vert{l}\rangle.
\end{multline}
In cold atom experiments, a wavepacket or atomic cloud can be boosted in momentum space in this way
using a `Bragg pulse'; this is commonly used to determine the excitation spectrum of cold atom
systems using Bragg spectroscopy (see, e.g., Ref.\ \cite{EGK10} for a description of the
experimental technique).  In experimental spectroscopy, both energy and momentum are well-resolved
in order to obtain the energy-momentum dispersion.  In our case, we supply the wavepacket with a
momentum using the factor $e^{i\mathbf{k}_0\cdot \mathbf{r}_l}$, but do not specify energy.  This
can be thought of as a Bragg pulse with sharp momentum resolution but poor or non-existent energy
resolution.  This allows us to explore various occupancies of the two energy bands.

The momentum zero wavepacket \eqref{eq:gaussian_wavepacket} turns out to predominantly have overlap
with eigenstates of $\mathcal{H}_{\text{HM}}$ at the bottom of the spectrum, in the lower band, as
long as $\sigma$ is not too small.  This is generally true in simple lattice models with negative
hopping constants.  In a complicated model like $\mathcal{H}_{\text{HM}}$, this is not \emph{a
  priori} obvious, but is the case for the parameters we are using.

\begin{figure}
 \includegraphics[width=0.99\linewidth]{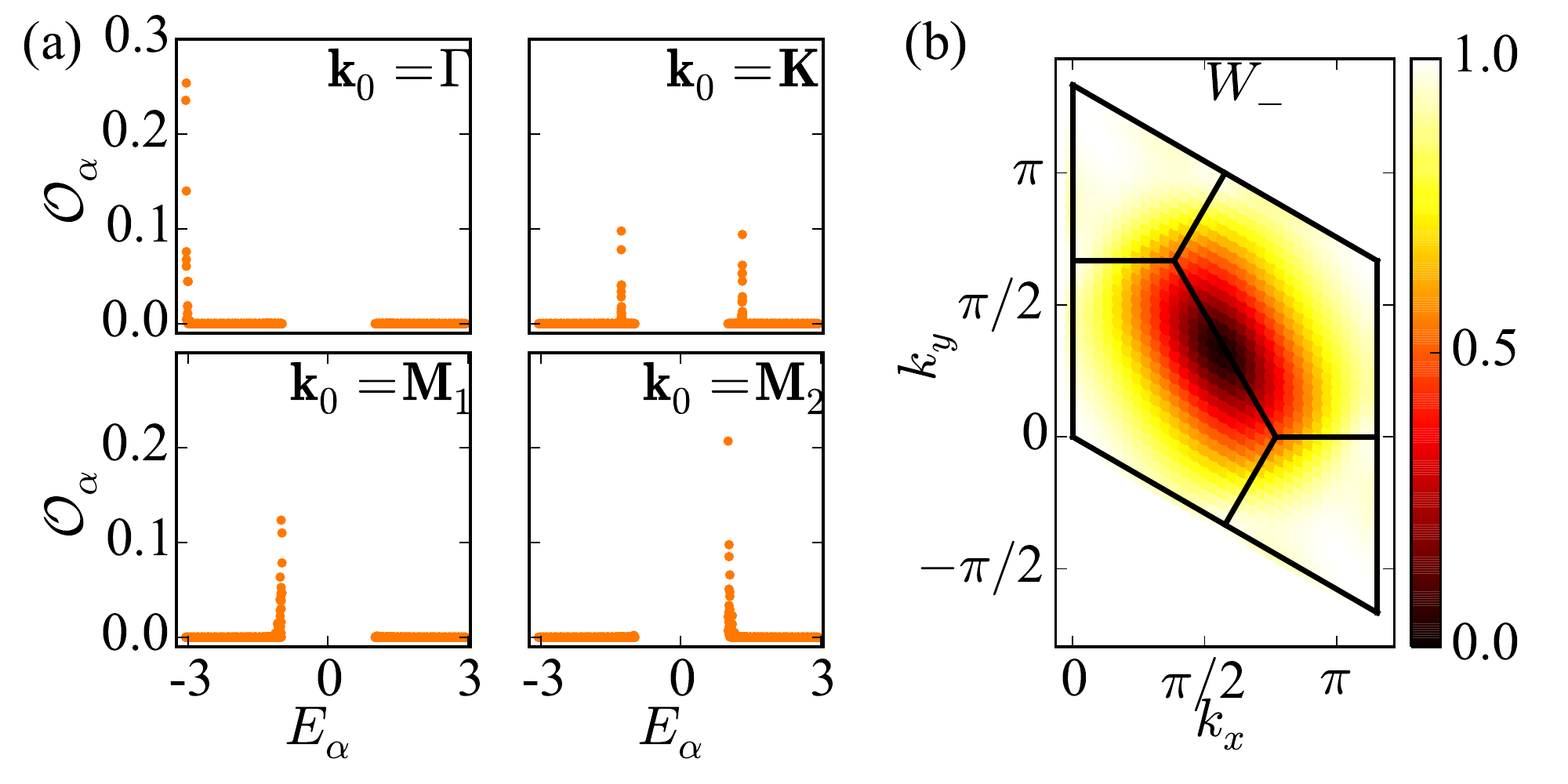}
 \caption{(Color online) (a) The overlaps $\mathcal{O}_\alpha$ of the initial wavepacket at different initial
   momenta $\mathbf{k_0}$ with the eigenstates of $\mathcal{H}_{\text{HM}}$ (with periodic boundary
   conditions) are plotted against the energy eigenvalues $E_\alpha$. (b) The weight of the
   wavepacket on the lower band ($W_-$) for each initial momentum in the Brillouin zone.}
 \label{fig:weight_lower}
\end{figure}

Boosting the wavepacket in momentum space as in Eq.\ \eqref{eq:initial_wavepacket_with_momentum} can
result in the wavepacket having support on both the bands of $\mathcal{H}_{\text{HM}}$.  This is
exemplified in Fig.~\ref{fig:weight_lower} through the overlap of the wavepacket with the
eigenstates of $\mathcal{H}_{\text{HM}}$. We denote the overlap of the initial state,
$\vert\psi(t=0)\rangle$ with an eigenstate of $\mathcal{H}_{\text{HM}}$, $\vert u_\alpha\rangle$
with eigenvalue $E_\alpha$ as $\mathcal{O}_{\alpha} = \vert\langle\psi(t=0)\vert
u_\alpha\rangle\vert^2$. Fig~\ref{fig:weight_lower}(a) shows a plot of $\mathcal{O}_\alpha$ against
$E_\alpha$ for four out of the six high-symmetry momentum points being the initial momentum of the
wavepacket. It can be seen that for $\mathbf{k}_0=\mathbf{K}$ the wavepacket has support on both
bands.  The wavepacket corresponding to $\mathbf{k}_0=\mathbf{M}_1$ is also shifted higher in
energy, though it has overlaps primarily with the states of the lower band. On the contrary, for
$\mathbf{k}_0=\mathbf{M}_2$ the weight shifts almost entirely to the upper band.
Such a drastic difference of behavior between the $\mathbf{M}_1$ and $\mathbf{M}_2$ points may seem
unexpected because they are related by symmetry. 
However, the eigenfunction structures are of course inequivalent, so the overlap distributions after a momentum boost cannot
be expected to be similar.

We quantify the weight of the wavepacket on the lower band, $W_-$ by taking the sum of the overlaps of the wavepacket with the eigenstates of the lower band. Mathematically, 
\begin{equation}
\label{eq:Wpm}
W_-=\sum_{\alpha=1}^{N/2}\mathcal{O}_\alpha; ~~~~~ W_+ = 1-W_-,
\end{equation}
where $W_+$ is the weight on the upper band, and $N$ is the number of sites in the lattice and hence
the number of single-particle eigenstates. The color map in Fig.~\ref{fig:weight_lower}(b) shows the magnitude of $W_-$ 
for a wavepacket with a given initial momentum in the Brillouin zone.  

\subsection{The trap}

The evolution in time is carried out with the Hamiltonian
\begin{equation}
\label{eq:Haldplustrap}
\ham ~=~ \ham_{\text{HM}} + \ham_{\text{trap}} ~=~ \ham_{\text{HM}} +  \sum_l V(l) \hat{b}_l^\dagger
\hat{b}_l ,
\end{equation}
where
\begin{equation}
\label{eq:trap}
V(l) = \dfrac{V_0}{2}\vert\mathbf{r}_{l}-\mathbf{r}_{0}\vert^2,
\end{equation}
with $\mathbf{r}_0=(x_0,y_0)$ denoting the center of the harmonic trap and $V_{0}$ controlling its
strength.

The force exerted by the trap is along the inward radial direction, so that one expects Bloch
oscillations in this direction.  We are particularly interested in the transverse response, and
hence the angular velocity of the wavepacket around the center of the trap is a natural observable
to study.  The angular variable $\theta$ at position $\mathbf{r}=(x,y)$ is given by $\theta =
\tan^{-1}\frac{y-y_0}{x-x_0}$. Its average as a function of time is calculated using
\begin{equation}
 \langle\theta\rangle(t) = \tan^{-1}\frac{\langle y\rangle(t) - y_0}{\langle x\rangle(t) - x_0},
\end{equation}
from the average $x$ and $y$ for the time dependent wavefunction
\begin{equation}
 \langle x \rangle (t) = \sum_l \vert c_l(t)\vert^2 x_l \, ,
\quad
 \langle y \rangle (t) = \sum_l \vert c_l(t)\vert^2 y_l \, .
\end{equation}
%
%

\subsection{Simulations of wavepacket dynamics}

\begin{figure*}[!ht]
 \includegraphics[width=0.99\linewidth]{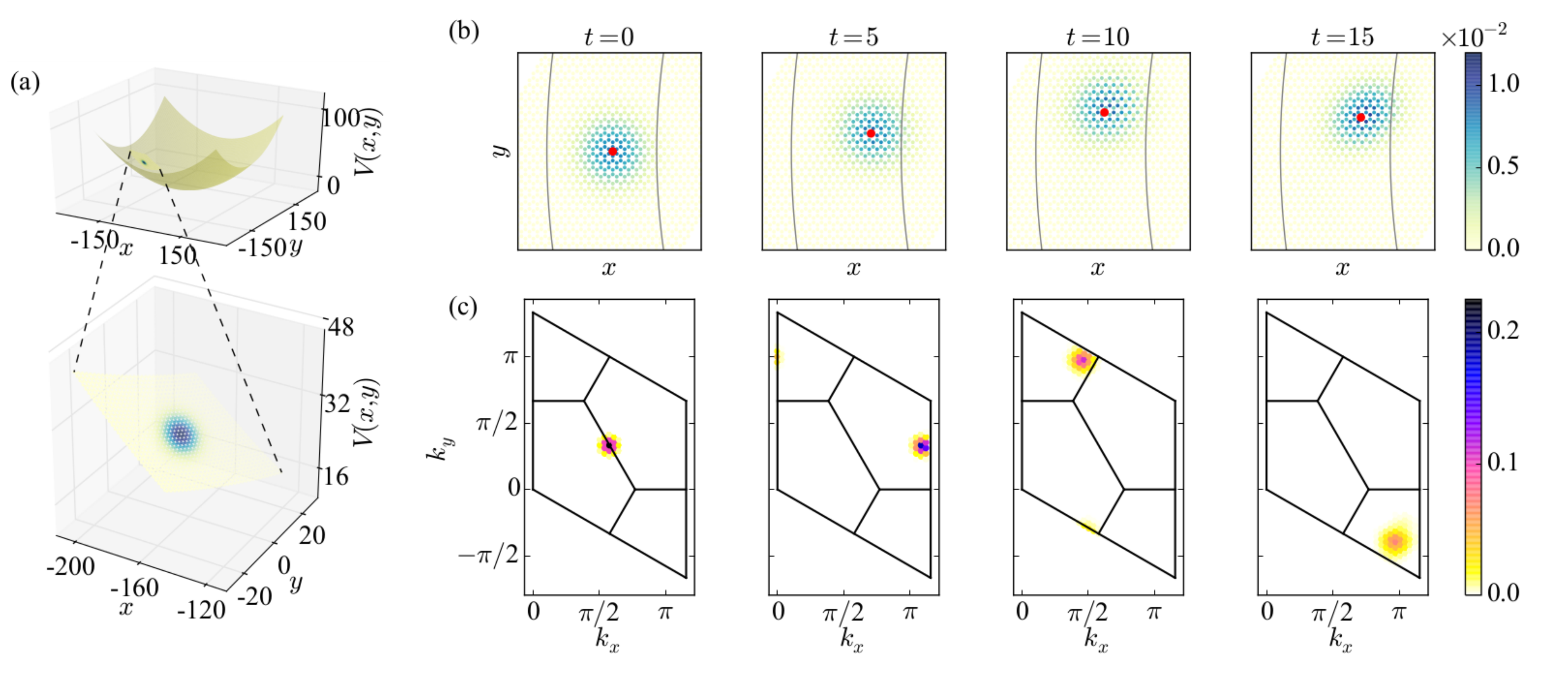}
 \caption{(Color online) (a) Geometry of the lattice and the trap. The surface represents the
   potential of the harmonic trap and the colors show the occupancies of the wavepacket at $t=0$ on
   the real space lattice. The lower figure shows a zoom near the wavepacket. (b) Snap shots of the
   time evolution of the real space occupancies ($\vert c_l(t)\vert^2$). The red circle shows the
   center of the mass of the wavepacket. The solid lines are equipotential contours of the trap
   potential. (c) Occupancies of each momentum mode in the Brillouin zone at different times ($\vert
   \tilde{c}_{\mathbf{k}}(t)\vert^2$).  The trap strength for all figures is $V_0=0.002$. The
   initial wavepacket has a gaussian width $\sigma^2=50$, initial momentum
   $\mathbf{k}_0=\mathbf{M}_2$, and $\vert\mathbf{r}_0-\mathbf{r}_c\vert=164.5$ in units of the
   lattice constant. }
 \label{fig:shots_r_offcenter}
\end{figure*}

In this work we present results for the exact dynamics of wavepackets placed off-center in the
trap, and compare with semiclassic predictions.

We present simulations mostly for a relatively weak trap ($V_0=0.002$), where the wavepacket
width $\sigma$ is much smaller than the distance ($\approx 164.5$) to the trap center
$\mathbf{r}_0$.  The trap potential gradient (i.e., the force) does not vary too much over the extent
of the wavepacket.   
In Sec.~\ref{sec:tight} we also present results for a tighter trap ($V_0=0.02$),
where the initial distance of the wavepacket to the trap center is $10$ times smaller so as to have
the same force at the center of the wavepacket.  The trap curvature is more significant in this
case.

In Fig.~\ref{fig:shots_r_offcenter}(b) we show the real-space evolution of the wavepacket in the
weak trap, for two different initial momenta.   
We focus on dynamics up to $t\approx20$.  The motion of the wavepacket on this timescale is not
large compared to its width.  The force acts in the radial (positive $x$) direction.  A
transverse response, perpendicular to the force, is clearly visible; we analyze this
quantitatively through the time dependence of $\langle\dot{\theta}\rangle$.
In the next sections we provide a thorough comparison of numerically exact results for
$\langle\dot{\theta}\rangle$ obtained through direct simulation (which we refer to as
$\langle\dot{\theta}\rangle_{\text{exact}}$) with predictions from the semiclassical formalism,
to be defined below.

\begin{figure}
 \includegraphics[width=0.99\linewidth]{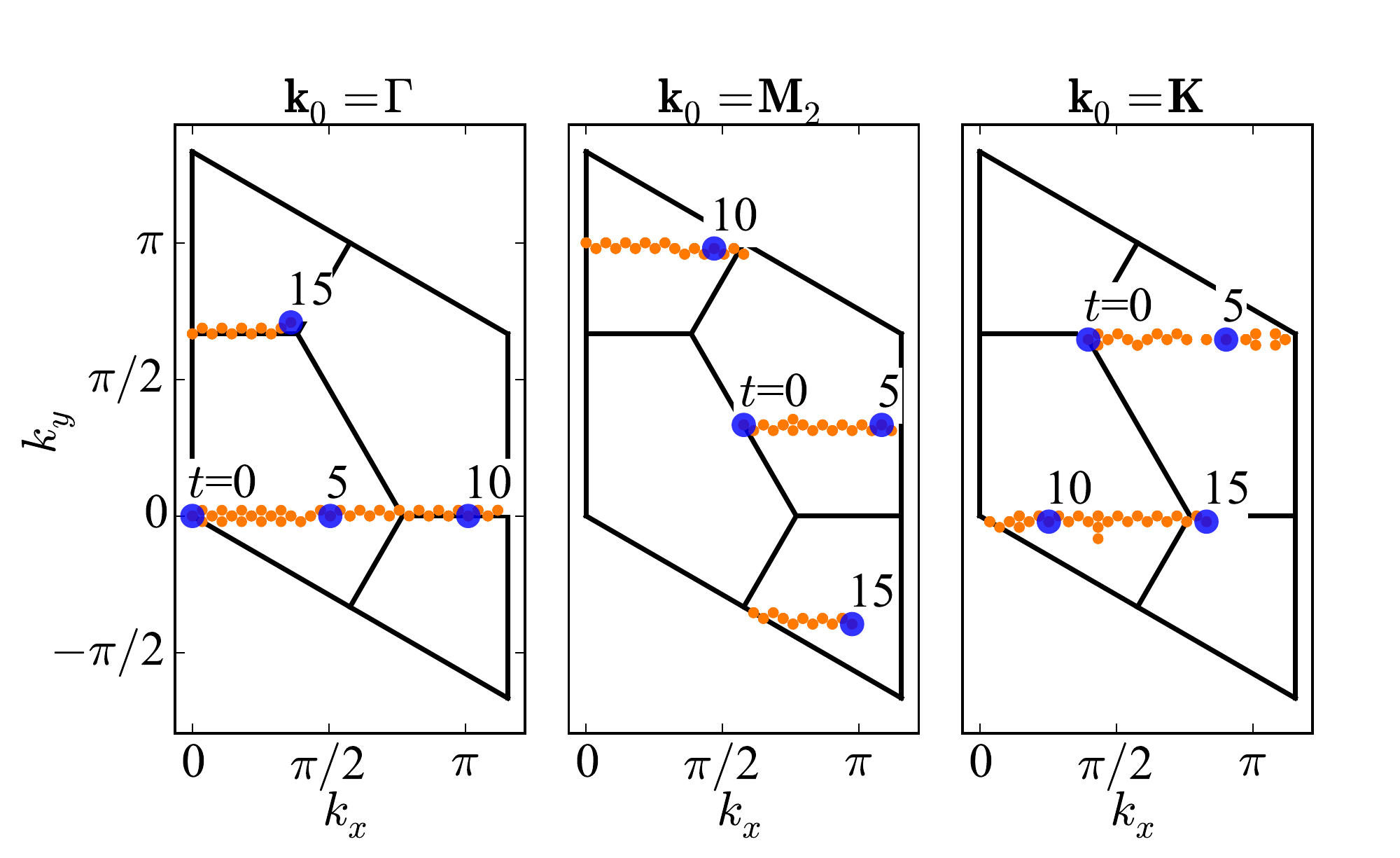}
 \caption{(Color online) Trajectories of the wavepacket in momentum space for three initial momenta
   showing Bloch oscillations.  The orange circles represent the trajectory of the center of the
   wavepacket in momentum space.  The blue larger circles denote the position of the 
   wavepacket center in momentum space at four instants of time ($t=0$, 5, 10, and 15).}
 \label{fig:traj}
\end{figure}

In addition to the transverse response, there are also Bloch oscillations in the radial direction;
this is not obvious in the real-space snapshots but is more evident in momentum space.
The motion of the wavepacket in momentum space is obtained by taking a lattice Fourier transform of
the coefficients $c_l(t)$ at each instant of time to obtain the occupancies of each momentum,
denoted by $\tilde{c}_{\mathbf{K}}(t)$. This motion can be visualized by plotting the coefficients
$\vert\tilde{c}_{\mathbf{k}}(t)\vert^2$ over the Brillouin zone at different instants of time, as
done in Fig.~\ref{fig:shots_r_offcenter}(c).  
The wavepacket moves through the Brillouin zone at constant rate in the direction of the force.  Due
to the periodicity of the Brillouin zone, each time the wavepacket exits through the right or top
boundary it re-enters through the left  or bottom boundary.  As a visual aid, we show
schematically in Fig.~\ref{fig:traj} the trajectories of the wavepacket centers starting from the
three high-symmetry points.

%

\section{Semiclassical dynamics \label{sec:semiclassics} }

In this section, we set up the semiclassical framework to calculate the time evolution of $\td$(t).
We first formulate  the basic `point-particle' approach, under the standard assumption of
simultaneously well-defined position and momentum.  We then formulate an extension where the
structure of the wavepacket in momentum space is taken into account.  

In the most basic semiclassical approach, the structure of the wavepacket in both real and momentum
space are ignored, so that the wavepacket or particle is described by a sharply defined position and
momentum $(\mathbf{r},\mathbf{k})$.  In addition, it is also assumed that the wavepacket dynamics is
completely governed by a single energy band.  We will first write down the semiclassical equations
with the assumption that the wavepacket has support only on the lower band.
The semiclassical equations of motion are~\cite{Xiao2010}
\begin{subequations} \label{eq:sc_botheqs}
\begin{equation}
 \frac{d\mathbf{r}}{dt} = \bm{\nabla}_{\mathbf{k}}E_-(\mathbf{k}) - \frac{d\mathbf{k}}{dt}\times\bm{\Omega}_-(\mathbf{k}),
 \label{eq:sc1}
\end{equation}
\begin{equation}
\frac{d\mathbf{k}}{dt} = \mathbf{F}.
 \label{eq:f}
\end{equation}
\end{subequations}
Here $E_-(\mathbf{k})$ is the energy dispersion, and $\bm{\Omega}_-(\mathbf{k})$ is the Berry
curvature of the lower band.  From the second term in Eq.~(\ref{eq:sc1}), we see that the Berry
curvature induces a velocity perpendicular to the direction of the external force, which leads to
the transverse motion of the wavepacket.

We now specialize to the geometry we are using, with a trapping potential centered at
($x_0$,$y_0$).  Using \eqref{eq:trap}, the external force is given by 
\begin{equation}
 \mathbf{F}(\mathbf{r}) = -\boldsymbol{\nabla}_{\mathbf{r}}V(\mathbf{r})=-V_0((x-x_0)\hat{x} +(y-y_0)\hat{y}).
\end{equation}
So, the semiclassical equations form a set of four coupled differential equations
\begin{subequations}
\begin{equation}
 v_{-,x}(\mathbf{k})=\frac{dx}{dt}=\frac{\partial E_-(\mathbf{k})}{\partial k_x} + V_0(y-y_0)\Omega_-^z(\mathbf{k}),
\end{equation}
\begin{equation}
 v_{-,y}(\mathbf{k}) = \frac{dy}{dt}=\frac{\partial E_-(\mathbf{k})}{\partial k_y} - V_0(x-x_0)\Omega_-^z(\mathbf{k}),
\end{equation}
\begin{equation}
 \frac{dk_x}{dt}=-V_0(x-x_0),
\end{equation}
\begin{equation}
 \frac{dk_y}{dt}=-V_0(y-y_0).
\end{equation}
\label{eq:scset}
\end{subequations}
This set of equations can be solved explicitly to trace out the trajectory in time of a particle in
real as well as momentum space.  We label as $\langle\dot\theta\rangle_{-,\text{pp-sc}}$ the angular
velocity corresponding to the real space trajectories calculated in this way. (The subscript `pp-sc'
stands for `point-particle semiclassics' and the $-$ sign denotes that the lower band properties
have been used).  A similar calculation can be done with the characteristics of the upper band
($E_+(\mathbf{k})$ and $\Omega_+^z(\mathbf{k})$) and the angular velocity so calculated is denoted
by $\langle\dot\theta\rangle_{+,\text{pp-sc}}$.  

As observed previously, wavepackets can have support on both bands.  One reasonable procedure would
be to use the $\langle\dot\theta(t)\rangle_{-,\text{pp-sc}}$ or
$\langle\dot\theta(t)\rangle_{+,\text{pp-sc}}$ curve, depending on whether the lower or upper band has
more occupancy.  We follow a somewhat more refined procedure by taking the weighted
average of the two according to the weights $W_{\mp}$ of the initial packet on the two bands. Hence the
angular velocity calculated from the point-particle semiclassics is defined as
\begin{equation}
\langle\dot\theta\rangle_{\text{pp-sc}} = W_-\langle\dot\theta\rangle_{-,\text{pp-sc}} + W_+\langle\dot\theta\rangle_{+,\text{pp-sc}}
\end{equation}

\begin{figure}[tbhp]
\begin{center}
 \includegraphics[width=0.9\columnwidth]{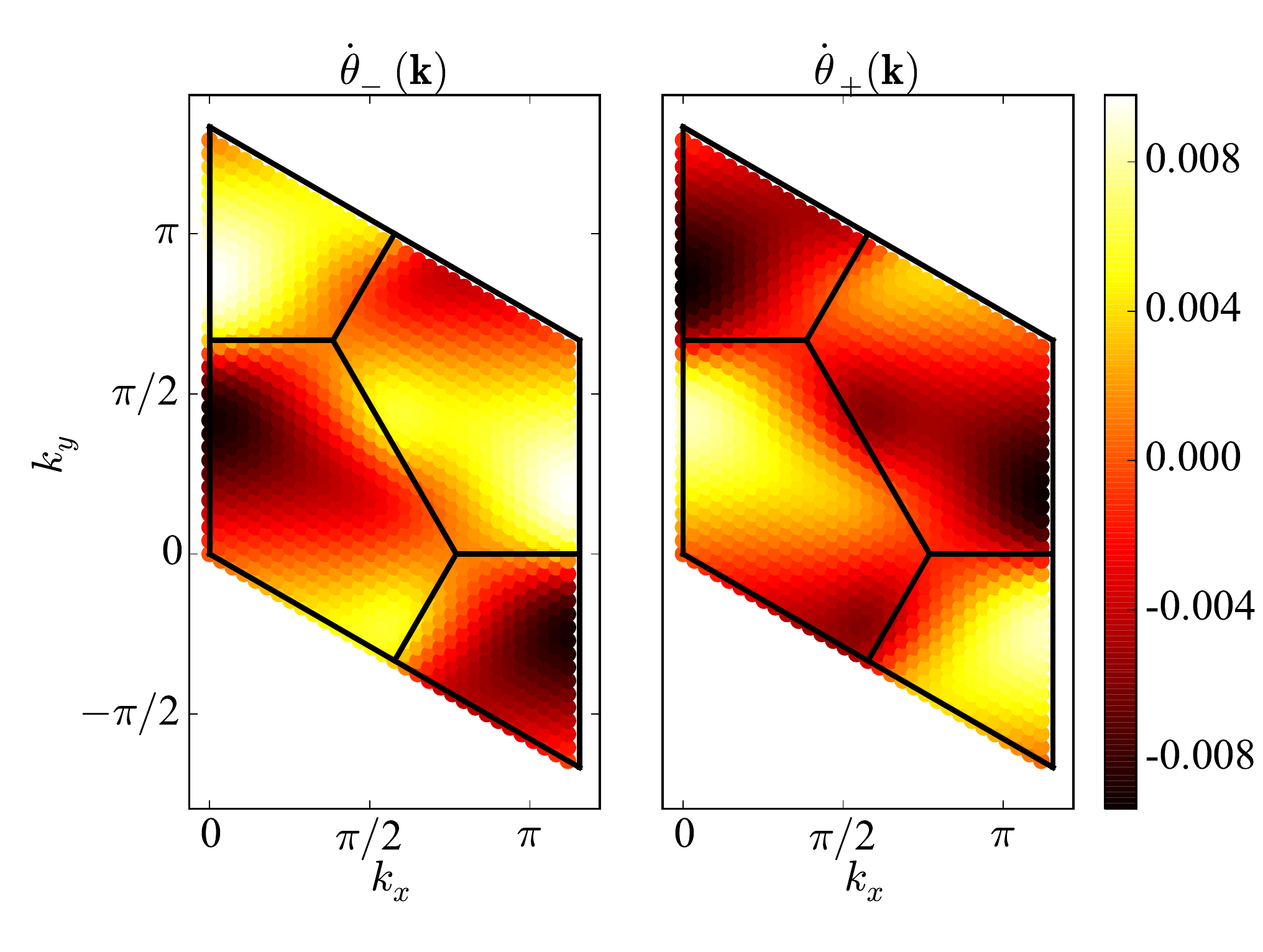}
 \end{center}
 \caption{(Color online) Profiles of $\dot\theta_-(\mathbf{k})$ and $\dot\theta_+(\mathbf{k})$ in
   the Brillouin zone as calculated from combining the semiclassical equations of motion
   Eq.~\eqref{eq:sc1} and the kinematic relation in Eq.~\eqref{eq:td}.  The coordinates of the
   center of the mass of wavepacket in real space relative to the center of the trap are given by
   $x-x_0=-164.5$ and $y-y_0=0.5$, and the trap strength is $V_0=0.002$.  
Here $\dot\theta_-(\mathbf{k})$ and $\dot\theta_+(\mathbf{k})$ have nearly
but not exactly equal and opposite values, $\dot\theta_-(\mathbf{k})\approx-\dot\theta_+(\mathbf{k})$.
}
 \label{fig:vx_vy_td}
\end{figure}

A key assumption above is that the wavepacket can be treated like a point particle in both real and
momentum space simultaneously, hence neglecting the quantum nature of the wavepacket. However in
realistic quantum experiments and simulations, where the wavepacket is of finite extent, the
validity of this assumption is not \emph{a priori} clear.
We now extend this formalism to take into account the finite spread of the wavepacket in momentum
space.
From the geometric definition $\theta = \tan^{-1}\frac{y-y_0}{x-x_0}$, we obtain
\begin{equation}
 \td =\frac{(x-x_0) v_y -(y-y_0)v_x}{(x-x_0)^2+(y-y_0)^2}.
 \label{eq:td}
\end{equation}
By using the expressions of $v_{\pm,x}(\mathbf{k})$ and $v_{\pm,y}(\mathbf{k})$ obtained from
Eq.~(\ref{eq:scset}a) and Eq.~(\ref{eq:scset}b), we define the functions
$\dot\theta_\pm(\mathbf{k})$ in the Brillouin zone.  Their typical profiles are shown in
Fig.~\ref{fig:vx_vy_td}, with parameter values corresponding to the initial position used in
Fig.~\ref{fig:shots_r_offcenter}.  

With our parameters, we have $E_-(\mathbf{k})\approx -E_+(\mathbf{k})$.  Also, we always have
$\Omega_+(\mathbf{k}) = -\Omega_-(\mathbf{k})$.  Hence we get $\dot\theta_-(\mathbf{k})\approx
-\dot\theta_+(\mathbf{k})$.  In other words the profiles shown in Fig.~\ref{fig:vx_vy_td} for the
two bands, $\dot\theta_{\pm}(\mathbf{k})$, are nearly but not exactly negative of each other.

These profiles of $\dot\theta_\pm(\mathbf{k})$ can be used to calculate the evolution of the angular
velocity in time by taking a weighted average of $\dot\theta_\pm(\mathbf{k})$, the weights being the
occupancies of the wavefunction in momentum space ($\vert\tilde{c}_{\mathbf{k}}(t)\vert^2$)
multiplied by the weights in each band ($W_\pm$) defined in \eqref{eq:Wpm}.  We denote the
angular velocity calculated this way as $\langle\dot\theta\rangle_{\text{wp-sc}}(t)$, the `wp' as a
reminder that the wavepacket structure is taken into account.  Thus
\begin{multline}
 \langle\td(t)\rangle_{\text{wp-sc}} = 
  W_- \langle\td(t)\rangle_{-,\text{wp-sc}} +   W_+ \langle\td\rangle_{+,\text{wp-sc}}
\\
=  W_- \sum_{\mathbf{k}\in \text{BZ}}\vert\tilde
 c_{\mathbf{k}}(t)\vert^2 \td_-(\mathbf{k})  ~+~  W_+ \sum_{\mathbf{k}\in \text{BZ}}\vert\tilde
 c_{\mathbf{k}}(t)\vert^2 \td_+(\mathbf{k})
\\
= \sum_{\mathbf{k}\in \text{BZ}}\vert\tilde{c}_{\mathbf{k}}(t)\vert^2
\left[ W_-\td_-(\mathbf{k})+W_+\td_+(\mathbf{k})\right]
\,.
\end{multline}
This procedure assumes that the force does not change along the spatial extent of the wavepacket;
the spread of the wavepacket in momentum space is taken into account while a point-particle
description is used in real space.  Therefore, this description will break down when the force
varies significantly within the real-space support region of the wavepacket (Sec.~\ref{sec:tight}).
In addition, note that this is not a computationally advantageous approximation for the time
evolution, since we are anyway solving the full problem in order to obtain the Fourier transform
$\tilde{c}_{\mathbf{k}}(t)$.  Our purpose here is to investigate whether (and how much) taking the
momentum-space spread into account improves the semiclassical description.

In this work we focus on parameter regimes such that the wavepacket does not have large
displacements in real space within the time scales $t\lesssim20$ of interest
(Fig.~\ref{fig:shots_r_offcenter}).  Therefore, we make a further simplifying assumption and take
$\td$ as position independent, setting $\mathbf{r}$ to be initial position of the wavepacket at
$t=0$, and use the resulting distribution of $\td$ to calculate the average.  In the following two
sections we test how well this procedure describes the angular motion of the
wavepacket.

%


 \section{Comparison of semiclassical predictions with exact dynamics \label{sec:thetadot}}

\begin{figure*}
 \includegraphics[width=0.95\linewidth]{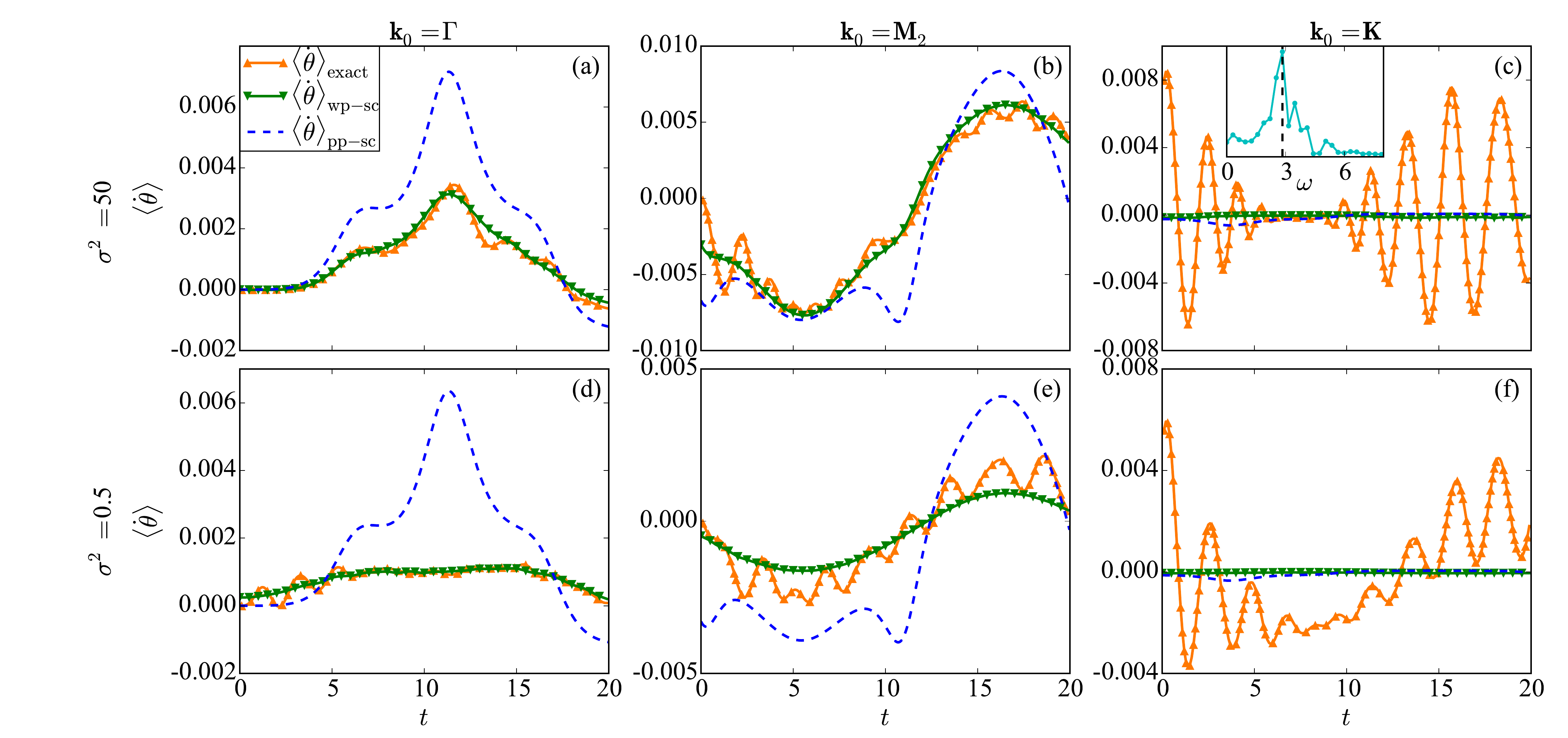}
 \caption{(Color online) 
The angular velocity from exact simulations  ($\langle\dot\theta\rangle_{\text{exact}}$), compared
with point-particle semiclassics  ($\langle\dot\theta\rangle_{\text{pp-sc}}$) and with extended
semiclassics taking momentum-space structure into account
($\langle\dot\theta\rangle_{\text{wp-sc}}$). 
Initial wavepacket size:  $\sigma^2=50$ [(a)-(c)] and $\sigma^2=0.5$ [(d)-(f)].  Three different
initial momenta are shown:  $\mathbf{k_0}=\mathbf{\Gamma}$ [(a),(d)], $\mathbf{k_0}=\mathbf{M_2}$
[(b),(e)], and $\mathbf{k_0}=\mathbf{K}$ [(c),(f)].  
For $\mathbf{k_0}=\mathbf{\Gamma}$ and $\mathbf{k_0}=\mathbf{M_2}$,
$\langle\dot\theta\rangle_{\text{wp-sc}}$ agrees quantitatively with
$\langle\dot\theta\rangle_{\text{exact}}$, whereas $\langle\dot\theta\rangle_{\text{pp-sc}}$ shows
qualitative agreement at best.  For $\mathbf{k}_0=\mathbf{K}$,
$\langle\dot\theta\rangle_{\text{exact}}$ shows oscillations which are not captured by semiclassics.
In panel (c), the Fourier transform of the oscillations in $\langle\dot\theta\rangle_{\text{exact}}$
is shown as inset. }
 \label{fig:comparisons}
\end{figure*}

In this section, we compare the angular velocities of the wavepacket obtained from the exact
simulations, ($\langle\dot\theta\rangle_{\text{exact}}$) to those obtained from the semiclassical
calculations ($\langle\dot\theta\rangle_{\text{pp-sc}}$ and
$\langle\dot\theta\rangle_{\text{wp-sc}}$) and discuss the regimes of validity of the
semiclassical framework.

In Fig.~\ref{fig:comparisons} we plot the angular velocities as a function of time for the setup
corresponding to that shown in Fig.~\ref{fig:shots_r_offcenter}(a) for two different sizes and three
different initial wavepacket momenta, $\mathbf{\Gamma}$, $\mathbf{M_2}$ and $\mathbf{K}$.
Before discussing in detail, we make some general observations:

\addtolength{\leftmargini}{-0.4\leftmargini}
\begin{itemize}
\addtolength{\itemsep}{-\itemsep}
\item For the $\mathbf{\Gamma}$ point and $\mathbf{M_2}$ point initial states,
  $\langle\dot\theta\rangle_{\text{pp-sc}}$, calculated using the basic `point-particle'
  semiclassical equations \eqref{eq:scset}, shows similar overall qualitative features as the
  evolution of the exact $\langle\dot\theta\rangle_{\text{exact}}$, but it generally fails to
  quantitatively reproduce the evolution.
On the other hand, $\langle\dot\theta\rangle_{\text{wp-sc}}$, calculated using the modified
semiclassics of Sec.~\ref{sec:semiclassics} (taking into account the wavepacket structure in
momentum space) reproduces many of the prominent features of the
$\langle\dot\theta(t)\rangle_{\text{exact}}$ curve.  

For the  $\mathbf{K}$ point initial state, there seems to be no noticeable agreement.
\item There is generally better agreement between the semiclassics and the exact evolution for the
  wavepacket that is larger in real space ($\sigma^2=50$, top row) compared to the smaller
  wavepacket ($\sigma^2=0.5$, bottom row).
\item For the larger wavepacket ($\sigma^2=50$, top row), the $\mathbf{\Gamma}$ point initial state
  is almost completely in the lower band ($W_-=0.999$) and the $\mathbf{M_2}$ point initial state is
  almost completely in the upper band ($W_+=0.985$).  Hence, using only the lower band
  ($\mathbf{\Gamma}$) or only the upper band ($\mathbf{M_2}$) would give very nearly the same
  semiclassical curves as the ones shown, which are $W_{\pm}$-weighted mixtures.

For the smaller wavepacket ($\sigma^2=0.5$, bottom row), the same is true with the
$\mathbf{\Gamma}$ point initial state ($W_-=0.99$), but the  $\mathbf{M_2}$ point initial state now
has significant contribution from the lower band as well ($W_+=0.74$).  This  leads to cancellation
of the weighted mixtures, so that, comparing panels (b) and (e), we see much smaller values of
$\langle\dot\theta\rangle$ for the smaller wavepacket. 

For the $\mathbf{K}$ point initial state, the contributions of the two bands largely cancel each
other, resulting in tiny semiclassical predictions for $\langle\dot\theta(t)\rangle$.
\end{itemize}

We now discuss in more detail the larger ($\sigma^2=50$) wavepacket (top row). A wavepacket with
zero initial momentum has support almost completely on the lower band ($W_-=0.999$), hence the
relevant profile of angular velocity is $\td_-(\mathbf{k})$. At the zero momentum
($\mathbf{\Gamma}$) point, both the gradient of the band dispersion and the Berry curvature in the
lower band are zero, leading to a zero angular velocity.  As a result, the wavepacket starts with
zero $\td$.  From the momentum space trajectory in Fig.~\ref{fig:traj}(a) and the Brillouin zone
profile of Fig.~\ref{fig:vx_vy_td}, one can infer that the wavepacket mostly moves through regions
of near-zero $\td$.  As a result, the $\td$ remains relatively small as seen in
Fig.~\ref{fig:comparisons}(a).  The momentum-space shape of the packet plays a strong role in this
case: as the trajectory lies roughly between positive and negative regions of $\td_-(\mathbf{k})$,
small variations of the shape can cause $\td$ to vary between positive and negative values.
Accordingly, the dynamics of $\langle\dot\theta(t)\rangle_{\text{exact}}$ is captured notably better
by the extended semiclassics $\langle\dot\theta\rangle_{\text{wp-sc}}$ than by the point-particle
approximation $\langle\dot\theta\rangle_{\text{pp-sc}}$.

As discussed in Sec.~\ref{sec:construction}, the wavepacket with initial momentum at $\mathbf{M}_2$
has support almost entirely on the upper band ($W_-=0.015$), hence the upper band characteristics
are more relevant here.  At the $\mathbf{M}_2$ point, although the gradient of the band dispersion
vanishes, the Berry curvature has a sharp peak [see Fig.~\ref{fig:model}(c)].  As a result, the
wavepacket gains a finite angular velocity almost immediately at $t\approx0$.  (Note that the
semiclassical approximations, by construction, start with nonzero $\td$ at $t=0$, which is the value
of $\td$ at the $\mathbf{M}_2$ point.  The physical or exact $\td$ starts at zero.)  Considering the
trajectory Fig.~\ref{fig:traj}(b) and the Brillouin zone profile of Fig.~\ref{fig:vx_vy_td}, we note
that the trajectory moves through regions of large $\td$; this is reflected in the larger absolute
values of Fig.~\ref{fig:comparisons}(b).
The trajectories in momentum space intersect regions of $\dot\theta_+(\mathbf{k})<0$ for $t\lesssim10$ to explore
regions $\dot\theta_+(\mathbf{k})>0$ at later times.  The change of sign can be seen in
Fig.~\ref{fig:comparisons}(b) in all three curves.

For the wavepacket with $\mathbf{k}_0=\mathbf{K}$, the dynamics of
$\langle\dot\theta\rangle_{\text{exact}}$ shows pronounced oscillations which preclude meaningful
comparison with the semiclassic predictions.  The oscillations are due to the fact that the initial
state created according to Eq.~\ref{eq:initial_wavepacket_with_momentum} has significant weight on
both lower and upper bands ($W_-=0.457$), with a relatively well-defined energy gap between
eigenstates occupied in the lower band and eigenstates occupied in the upper band.  This is seen
through the overlaps plotted in Fig.~\ref{fig:weight_lower}(a).  A Fourier transform of the
$\langle\dot{\theta}(t)\rangle_{\text{exact}}$, inset to Fig.~\ref{fig:comparisons}(c), shows that
the dominant frequency (peak around $\approx2.8$ with width $\approx0.3$) matches the energy
difference ($\approx2.6$) between eigenstates of high overlap, Fig.~\ref{fig:weight_lower}(a) top
right.  The weighted band averages shown as semiclassical predictions stay near zero, which could
be thought of as the value around which the exact $\langle\dot{\theta}(t)\rangle_{\text{exact}}$
oscillates, but it is currently unclear whether this is a coincidence.  It is also currently unclear
whether a more sophisticated way of incorporating multiple bands might allow the semiclassics to
reproduce the oscillatory behavior or the average curve around which
$\langle\dot{\theta}(t)\rangle_{\text{exact}}$ oscillates.

\begin{figure}
 \includegraphics[width=0.99\columnwidth]{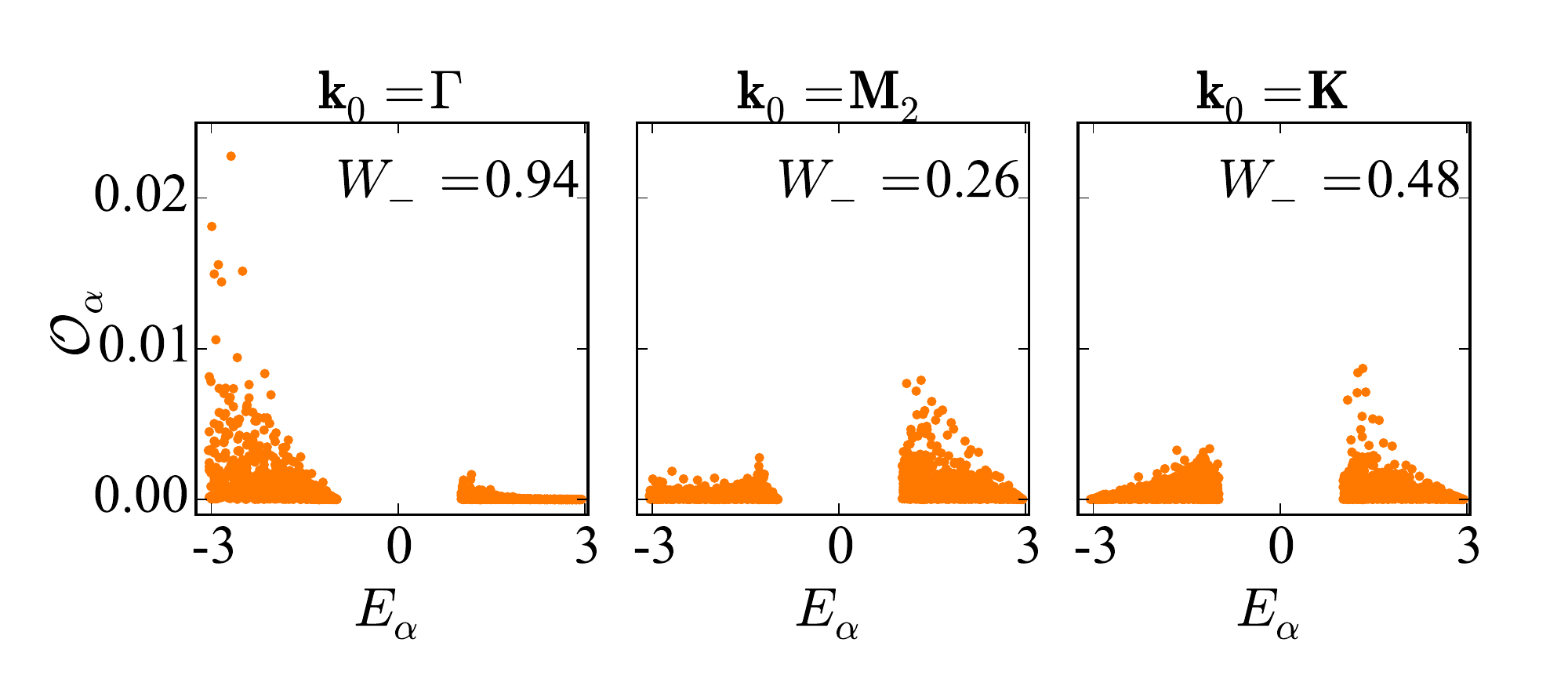}
 \caption{(Color online) 
   Overlaps as in Fig.\ \ref{fig:weight_lower}(a), for a smaller wavepacket, $\sigma^2=0.5$.
   Comparing with the case of $\sigma^2=50$ [Fig.~\ref{fig:weight_lower}(a)], this smaller
   wavepacket has a much more spread-out distribution of weights on the eigenstates of
   $\mathcal{H}_{\text{HM}}$. The weight of the wavepackets in the lower band $W_-$ (provided with
   each panel), are all closer to $1/2$ compared to the larger wavepacket of
   Fig.~\ref{fig:weight_lower}, where we had $W_-=0.999$, $0.015$, $0.457$ for these three momenta.
 }
 \label{fig:overlaps}
\end{figure}

In the lower panels of Fig.~\ref{fig:comparisons}, we have used an initial wavepacket with
$\sigma^2=0.5$.  The Gaussian wavepacket is centered at the center of a hexagon in real space, so
that even with such a small $\sigma$ there are six sites equally occupied.   The exact
$\langle\dot{\theta}(t)\rangle_{\text{exact}}$ now deviates significantly from the point-particle
semiclassics,  $\langle\dot\theta\rangle_{\text{pp-sc}}$.  The extended semiclassics,
$\langle\dot\theta\rangle_{\text{wp-sc}}$, continues to describe the overall behavior, for the
$\mathbf{\Gamma}$ point and $\mathbf{M_2}$ point initial states.  The exact dynamics now shows
oscillations for all three initial momenta.  This can be understood through the overlap
distribution, shown in Fig.~\ref{fig:overlaps}.  For the smaller packet, the overlaps are spread out
more in energy and also are far more biased toward more equal occupancies of the two bands ($W_-$
values are closer to $\frac{1}{2}$ compared to the corresponding values for the bigger packet).  As
a result, interference oscillations are visible also for the $\mathbf{\Gamma}$ point and
$\mathbf{M_2}$ point initial states, panels (d) and (e).

\section{Semiclassics in a tight trap \label{sec:tight}}

\begin{figure*}
 \includegraphics[width=0.95\linewidth]{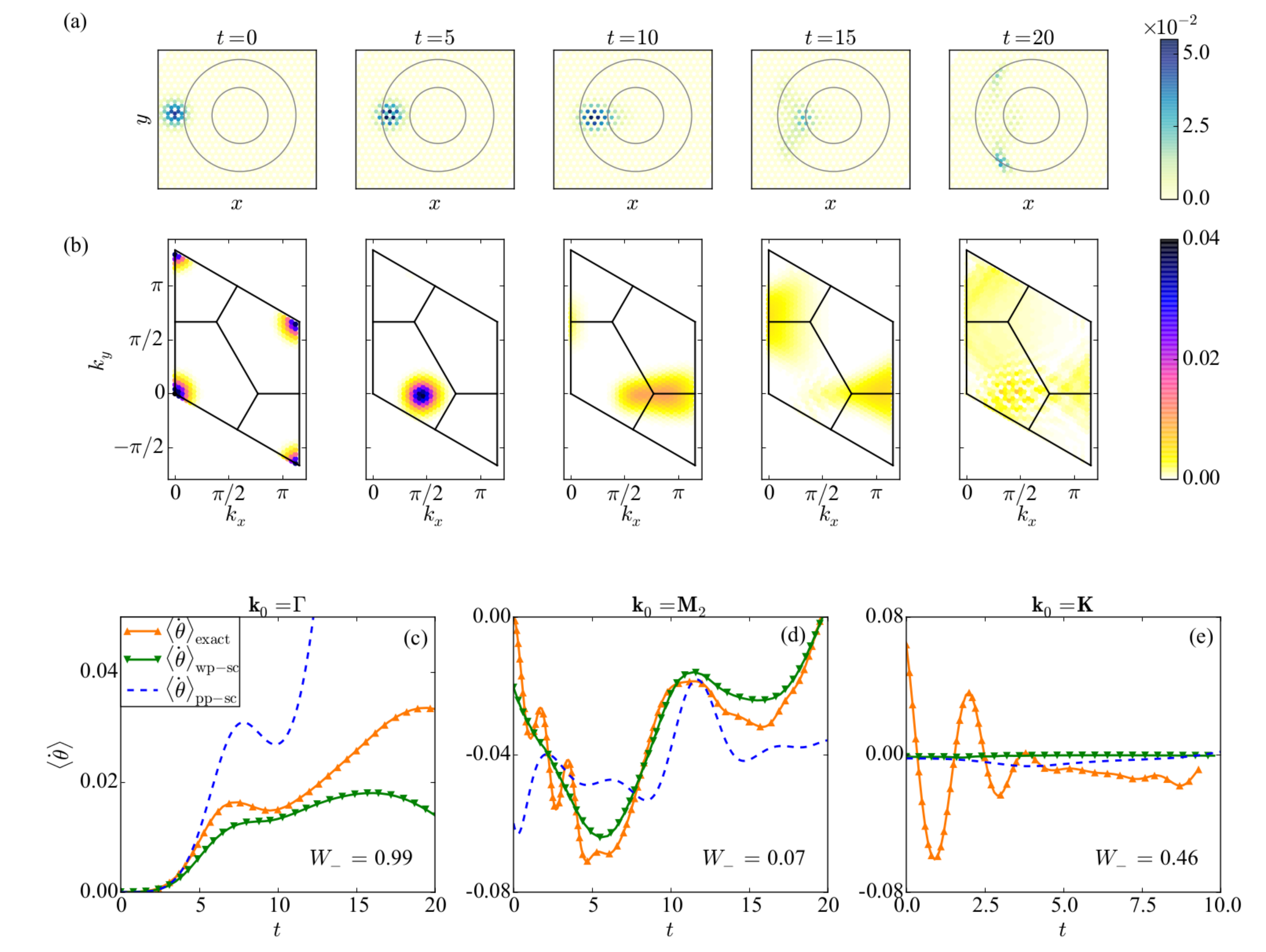}\\
 \caption{(Color online) Dynamics in a tight trap, $V_0=0.02$, initial position $\vert\mathbf{r}_0 -
   \mathbf{r}_c\vert=16.45$, initial size: $\sigma^2\approx10$.  (a,b) Snapshots for
   $\mathbf{k}_0=\mathbf{\Gamma}$.  (a) Real-space occupancies ($\vert c_l(t)\vert^2$).  Solid lines
   are equipotential contours of the trap potential. (b) Corresponding momentum space occupancies
   $\vert\tilde{c}_{\mathbf{k}}(t)\vert^2$ in the Brillouin zone.  (c) Comparison between the
   angular velocities calculated from full simulations ($\langle\dot\theta\rangle_{\text{exact}}$),
   with the two types of semiclassics ($\langle\dot\theta\rangle_{\text{pp-sc}}$ and
   $\langle\dot\theta\rangle_{\text{wp-sc}}$).}
 \label{fig:comparisons_tight}
\end{figure*}

In Sec.~\ref{sec:thetadot} we showed that, as the spread of the initial wavepacket in real space is
made bigger, the agreement between the results from the semiclassical approximations and exact
simulations improves.  The idea is that, increasing size in real space corresponds to decreasing
size in momentum space (as reflected also in decreasing spread in energy space); thus the
point-particle approximation in momentum space is more justified.
However, increasing real-space size can also lead to violation of semiclassics, as the semiclassic
equations of motion also assume sharply defined position.  One effect is that the finite spread of
the wavepacket in real space could lead to different potential gradients (different forces) at
different points within the wave packet.  This effect would not play a role for a constant gradient
but can occur in a harmonic trap.
This kind of `tidal' force, makes the point particle notion less justified in real space.

In order to characterize this effect, we consider the geometry of
Fig.~\ref{fig:comparisons_tight}(a) with $V_0=0.02$, one order of magnitude stronger than that in
Fig.~\ref{fig:shots_r_offcenter}.
The distance between the center of the trap and the center of the wavepacket is adjusted such that
the force at the center of the wavepacket remains the same compared to the geometry shown in Fig.~\ref{fig:shots_r_offcenter}(a).
It can be seen from Fig.~\ref{fig:comparisons_tight}(a) that, in the course of time evolution, the
wavepacket breaks apart, spreads out and does not keep the notion of a well defined wavepacket as
compared to the shallower trap [Fig.~\ref{fig:shots_r_offcenter}].
This is also reflected in the evolution of the wavepacket in momentum space as shown in Fig.~\ref{fig:comparisons_tight}(b).
For similar time scales as those studied in previous sections, the wavepacket in momentum space gets distorted and diffuses out completely, unlike the case in 
Fig.~\ref{fig:shots_r_offcenter}(c) where there is still a notion of a well defined peak centered around some value of momentum.
Quite surprisingly, our semiclassical calculation does not seem to fail completely even in this
extreme case since $\langle\dot\theta\rangle_{\text{wp-sc}}$ and
$\langle\dot\theta\rangle_{\text{exact}}$ still agree qualitatively [see Fig.\ref{fig:comparisons_tight}].
Even the point-particle semiclassics,  $\langle\dot\theta\rangle_{\text{pp-sc}}$, reproduces
qualitatively some of the peaks and dips of the exact curve. 

As in previous cases, for  $\mathbf{k}_0=\mathbf{K}$ there are  strong oscillations  due to
occupancies in both bands.

\section{Summary and Discussion \label{sec:summary}}

In this work, we have explored the dynamics of a gaussian wavepacket, with and without initial
crystal momentum, on the Haldane honeycomb Chern-band lattice in the presence of external forces
provided by a harmonic potential.  We have focused on short-time dynamics and compared to
semiclassical treatments.  Semiclassical descriptions of wavepacket dynamics are obviously
appealing, but the range of applicability is poorly explored.  This work can be regarded as a step
toward obtaining detailed intuition on the regimes of validity of the semiclassical manner of
predicting trajectories.  We have formally treated single-particle dynamics, but our considerations
should be applicable to non-interacting Bose condensates, and approximately to weakly interacting
Bose condensates.

We have found that the point-particle semiclassics reproduces many qualitative features even for
wavefunctions that are quite small in real space, and hence extended over non-negligible portions of
the Brillouin zone.  When this momentum-space extent is taken into account, the agreement can be
excellent even when the point-particle approximation fails.  This shows that the basic idea of
semiclassics, following position and momentum simultaneously, can correctly embody quantum dynamics
even when the point-particle approach fails.  Using a tight trap, we have also shown that this
extended semiclassical approach can function reasonably even when the wavepacket is completely
distorted or even torn apart (Fig.~\ref{fig:comparisons_tight}).  While this is reassuring for the
philosophy behind semiclassics, it does not immediately lead to a computationally advantageous
approximation, since to obtain the momentum-space structure we first evolved the complete system in
time (i.e. solved the problem computationally).  However, one can envision an extended semiclassics
where the wavepacket in momentum space is assumed to have fixed shape and the center moves according
to the point-particle equations \eqref{eq:sc_botheqs}.

This study is directly motivated by recent developments such as the experimental realization of
Haldane's honeycomb model in a cold-atom system \cite{Jotzu2014}, the experimental and theoretical
interest in the response of a localized wavepacket to an applied force (potential gradient)
\cite{Price2012,Dauphin2013,Jotzu2014, Aidelsburger2015,Duca2015}, and recent theoretical studies of
dynamics in various backgrounds using semiclassics \cite{Diener2003,Price2012,Price2013,Pruneda2009,
  Culcer2005,Wickles2013,Pettini2011, Zhang2010}.  In 2D lattices, a potential gradient will of
course lead to Bloch oscillations in the gradient direction (as widely studied, e.g., in
\cite{Kolovsky2003,Witthaut2004,Mossmann2005,Trompeter2006, Mossmann2005,Zhang2010, Tarruell2012,
  Uehlinger2013a}), but may also induce a transverse response.  This can happen even without Berry
curvature \cite{Zhang2010}, simply due to the structure of the energy band.  In the presence of
Berry curvature, the transverse response occurs due to a combination of the two effects.

The present study opens up many new questions.  First, we have focused on timescales such that the
wavepacket displacements are of the order of the wavepacket size.  It remains an open question to
see how well semiclassics works when the trajectories are long compared to the wavepacket size.  For
example, Refs.\ \cite{Diener2003,Price2012} have predicted a dramatic turning point in the
trajectory of a wavepacket traveling through a honeycomb (graphene) lattice with a constant
potential gradient.  It is unclear how closely a finite-sized quantum wavepacket would follow such a
prediction, especially when the position dynamics couples to internal distortion dynamics of a
realistic wavepacket.  Second, it remains an open question whether some version of our ``extended
semiclassics'' can be fashioned into a computationally useful approximation scheme.  Third, our
treatment of multiple-band occupancies is rather primitive (an incoherent average), and is unable to
account for interference oscillations.  Clearly, development of multiple-band semiclassics for such
time-evolution phenomena, perhaps along the lines of Ref.\ \cite{Culcer2005}, is called for.
Finally, since cold-atom experiments are more likely to track interacting Bose condensates or
fermionic clouds rather than single-particle wavepackets, it is of interest to find out in which
situations the dynamics of many-fermion and many-boson clouds resemble single-particle wavepacket
dynamics.


\section{acknowledgments}

MH thanks A.~Eckardt and C.~Gro{\ss} for useful discussions.

\appendix

\section{Momentum space properties of the Haldane model \label{sec:berrycurvature}}

The semiclassical equations of motion \eqref{eq:sc_botheqs} take the gradient of the energy
dispersion and the Berry curvature in momentum space as input. In this Appendix, we give some
details of these momentum space properties of the Haldane model, for completeness. Being realized on
a  lattice with a two-site basis, the Haldane model Hamiltonian in momentum space can be written as a collection
of $2\times2$ Hamiltonians $h(\mathbf{k})$:
\begin{equation}
 \ham_{\text{HM}} = \sum_{\mathbf{k}\in BZ}\Psi_\mathbf{k}^\dagger ~ h(\mathbf{k}) ~ \Psi_\mathbf{k},
 \end{equation}
 with
 \begin{equation}
 h(\mathbf{k})= \sigma^0 B^0({\mathbf{k}}) + \bm{\sigma}\cdot\mathbf{B}({\mathbf{k}})
 \label{eq:hamk}
\end{equation}
where $\Psi_\mathbf{k}^\dagger = (\hat{b}_{A,\mathbf{k}}^\dagger,\hat{b}_{B,\mathbf{k}}^\dagger)$
and $\hat{b}_{A(B),\mathbf{k}}^\dagger$ creates a Bloch state with momentum $\mathbf{k}$ on
sublattice $A(B)$. The $\sigma$'s are Pauli matrices with $\sigma^0$ being the identity matrix.  We
label the vectors connecting the nearest-neighbor sites as $\mathbf{a_1}=(0,-1),$
$\mathbf{a}_2=(\sqrt{3}/2,1/2)$, and $\mathbf{a}_3 = (-\sqrt{3}/2,1/2)$, and the vectors connecting
the next-nearest-neighbors as $\mathbf{b}_1 = \mathbf{a}_2-\mathbf{a}_3$, $\mathbf{b}_2 =
\mathbf{a}_3-\mathbf{a}_1$, and $\mathbf{b}_3 = \mathbf{a}_1-\mathbf{a}_2$.
With these notations, $B_0$ and $\mathbf{B}$ turn out to be
\begin{subequations}
 \begin{equation}
  B_0(\mathbf{k}) = 2J_2\cos\phi\sum_{i=1}^3\cos\mathbf{k}\cdot\mathbf{b}_i,
 \end{equation}
  \begin{equation}
  B_x(\mathbf{k}) = J_1\sum_{i=1}^3\cos\mathbf{k}\cdot\mathbf{a}_i,
 \end{equation}
 \begin{equation}
  B_y(\mathbf{k}) = J_1\sum_{i=1}^3\sin\mathbf{k}\cdot\mathbf{a}_i,
 \end{equation}
\begin{equation}
  B_z(\mathbf{k}) = -2J_2\sin\phi\sum_{i=1}^3\sin\mathbf{k}\cdot\mathbf{b}_i.
 \end{equation}
\end{subequations}
The energy dispersions of the two bands are 
\begin{equation}
E_\pm(\mathbf{k}) = B_0(\mathbf{k}) \pm \vert \mathbf{B}(\mathbf{k})\vert^2,
\end{equation}
where $+(-)$ denote the upper (lower) band.

The eigenstates of the Hamiltonian \eqref{eq:hamk} for a given momentum $\mathbf{k}$ can be written as 
\begin{equation}
 u_{+,\mathbf{k}}=\begin{pmatrix}e^{-i\frac{\zeta_{\mathbf{k}}}{2}}\cos\frac{\eta_\mathbf{k}}{2}\\e^{+i\frac{\zeta_{\mathbf{k}}}{2}}\sin\frac{\eta_\mathbf{k}}{2}\end{pmatrix};~u_{-,\mathbf{k}}=\begin{pmatrix}e^{-i\frac{\zeta_{\mathbf{k}}}{2}}\sin\frac{\eta_\mathbf{k}}{2}\\-e^{+i\frac{\zeta_{\mathbf{k}}}{2}}\cos\frac{\eta_\mathbf{k}}{2}\end{pmatrix},
\end{equation}
where $\eta_{\mathbf{k}}$ and $\zeta_{\mathbf{k}}$ are defined via
\begin{equation}
 \eta_{\mathbf{k}} = \cos^{-1}\frac{B_z(\mathbf{k})}{\vert\mathbf{B}(\mathbf{k})\vert};~~ \zeta_{\mathbf{k}}=\tan^{-1}\frac{B_y(\mathbf{k})}{B_x(\mathbf{k})}.
\end{equation}
With these notations, the Berry curvature is given by
\begin{equation}
 \Omega_\pm=\mp \frac{1}{4\pi}\bm{\epsilon}_{\mu\nu}[\partial_{k_\mu}\cos\eta_{\mathbf{k}}][\partial_{k_\nu}\zeta_{\mathbf{k}}]
\end{equation}
where $\bm{\epsilon}_{\mu\nu}$ is the two-dimensional Levi-Civita and $\Omega_{-(+)}$ refers to the Berry curvature of the lower (upper) band.

\bibliographystyle{apsrev4-1.bst}
\bibliography{references}

\end{document}